\documentclass[aps,pra,twocolumn,groupedaddress]{revtex4-2}

\usepackage{graphicx}

\begin{document}

\title{Erasing the which-path information of photons}

\author{Jinxian Guo$^{1,4}$, Qizhang Yuan$^{2,}$}
\email{lphysics@shnu.edu.cn}
\author{Yuan Wu$^{3}$, Weiping Zhang$^{1,4,5}$}
\affiliation{$^{1}$School of Physics and Astronomy, and Tsung-Dao Lee Institute, Shanghai
Jiao Tong University, Shanghai 200240, P. R. China}
\affiliation{$^{2}$Department of Physics, Mathematics and Science College of Shanghai Normal University, Shanghai 200234, P. R. China}
\affiliation{$^{3}$State Key Laboratory of Precision Spectroscopy, Quantum Institute
of Atom and Light, Department of Physics, East China Normal University,
Shanghai 200062, P. R. China}
\affiliation{$^{4}$Shanghai Research Center for
Quantum Sciences, Shanghai 201315, P. R. China}
\affiliation{$^{5}$Collaborative Innovation Center of Extreme Optics, Shanxi University,
Taiyuan, Shanxi 030006, P. R. China}
\date{\today}

\begin{abstract}
Which-path information of a quantum particle in interferometers is the key to infer the past of quantum particle. It arises many extensive discussions including quantum complementarity and path-visibility relation. The basic of these discussions are the description, detection and control of which-path information. In this article, we focus on the investigation of multidimensional which-path information in nested Mach-Zehnder interferometer. A general expression of which-path information is given and can be partially extracted by different detection method. Further analysis shows that the which-path information can be controlled by the phase differences and beam splitting ratios between the arms of nested Mach-Zehnder interferometer. Moreover, a new which-path information elimination phenomenon has been predicted and demonstrated experimentally. Our work can help to understand the physics of quantum particles, potentially apply to quantum information process and quantum metrology.
\end{abstract}

\maketitle

\section{Introduction}

Which-path information (WPI) is the information one has about which path a quantum particle took through a device \cite{wpi1,Jeff}. By measuring it in a detector, WPI has been used to infer the past of a quantum particle \cite{Walls, Rempe, Poto, Robert, Yuji}. For example, Wheler et al. \cite{Wheeler, Wheeler2, science2} found a counter-factual result: we can't infer the past of photons, because photons can hide their past. Up to now, WPI is still the key of discussions about basic quantum principles\cite{Scully, Leggett, Walls, Zeilinger, Rempe, Shi, wpi1, wpi2, wpi3}, including quantum complementarity and path-visibility relation. These discussions are often accompanied by the elimination of WPI in various interferometers, including double slit interferometer \cite{Shi, Wheeler, Scully, Walls, Leggett} and Mach-Zehnder interferometer (MZI) \cite{Science, Vaidman1, Vaidman2, Vaidman3, Nikolaev, Hellmuth, Khoon, Alonso, Sokolovski, Danan, Yuan2}. 

Recently, Danan et. al. find that WPI can be discontinuous \cite{Danan} in nested Mach-Zehnder interferometer (NMZI), which arises extensive discussions \cite{Vaidman1, Vaidman2, Vaidman3, Vaidman4, Vaidman5, Vaidman6, Vaidman7, Suhail1, Suhail2, Suhail3, Yuji, Robert, Poto, Nikolaev, Alonso, Yuan1, Yuan2}. In these discussions, several WPI elimination phenomena \cite{Zeilinger, Alonso, Yuan1, Yuan2} have been pointed out which highly depend on the state of the interferometer, including the stability and relative phases of the interferometer. Are there any other WPI elimination phenomena? Moreover, the detection method used in NMZI is a position sensitive one, totally different from Wheeler's case. Is WPI elimination in NMZI related to the detection method? To further understand WPI elimination phenomenon and the underlying physical mechanism, the description, detection and control of WPI need to be further investigated in NMZI.

In this article, we use a three-path interference model and focus on the WPI in NMZI. With the vibrations of mirrors in our interferometer, the position and phase information of photons have been changed simultaneously. A general expression of WPI can be obtained which shows multidimensional informations. Surprisingly, such multidimensional WPI can partially extracted by different detection methods. As examples, we use position sensitive detection and phase sensitive detection respectively to extract the position and phase induced WPI. Correspondingly, elimination condition of WPI is reconsidered under certain detection method. A detail discussion shows that the WPI can be controlled by the phase differences between interference arms and the splitting ratios of beam splitters (BSs) in NMZI. These results not only explain the experimental results in Ref. \cite{Danan}, but also predict a new WPI elimination phenomenon which has been demonstrated experimentally in our article.

This paper is organized as follows. In Sec. II, a three-path interferometer model is presented with adjustable splitting ratios of the beam splitters, phase differences between arms and multiple detection methods. A detail analysis and discussion of WPI is in Sec. III. Then we establish an experiment setup and verify the theoretical prediction of WPI elimination condition in Sec. IV. The conclusion is given in Sec. V.

\section{Theoretical model}

\begin{figure}[ht!]
\centering
\includegraphics[scale = 0.6]{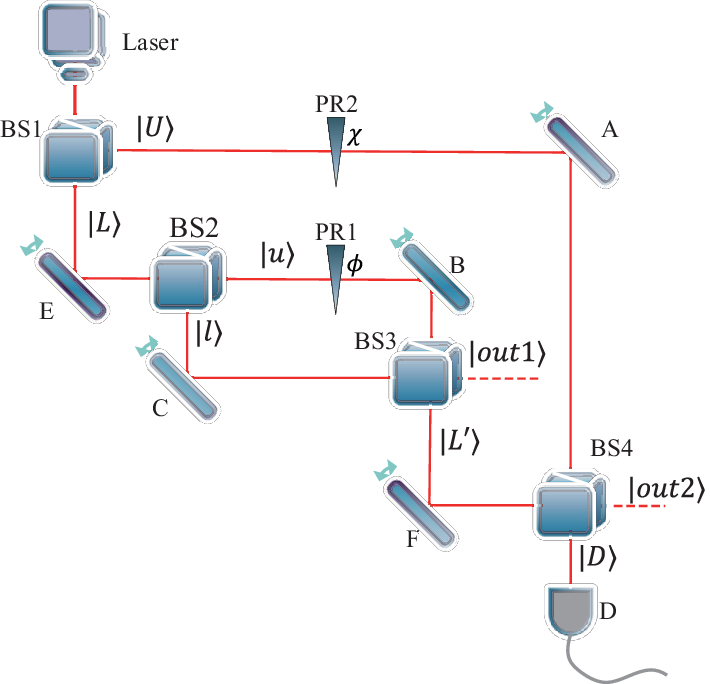}
\caption{Scheme of NMZI. All the splitting ratios of the beam splitters (BSs) used in this NMZI are adjustable. The mirrors A, B, C, E, F are vibrating with different frequencies $f_{k}, k=a,b,c,e,f$. PR1, PR2 represent the phase retarders and D is the photon detector.\label{fig1}}
\end{figure}

Our model is shown in the Fig. \ref{fig1}, photons are emitted from the laser source and enter the interferometer. Different from the previous work, the interferometer is built up by BSs with adjustable beam-splitting ratios. The mirrors in interferometer \cite{Suhail1, Suhail2, Suhail3, Yuji, Robert, Poto, Nikolaev, Alonso, Yuan1, Yuan2} oscillate slightly with distinguishable frequencies $f_{k}$ which introduce additional complex phase shifts $\varphi_{k}$ where $k=a,b,c,e,f$. The real part and imaginary part of $\varphi_{k}$ represent the change of the optical phase and position of light beam caused by the vibration\cite{Yuan1, Yuan2}, respectively. The WPI of photons is indicated by $\varphi_{k}$ which can be extracted in the detector. For example, if we recognize $\varphi_{a}$ in the detector, we know that the photons have been reflected by mirror A. To control the WPI, we introduce two phase shifts $\phi$ and $\chi$ by the tunable phase retarders (PR1, PR2 showing in Fig. \ref{fig1}) when photons pass through the interferometer in three different paths. 

In detail, when photons emit from the light source, the photon state passing through the BS1 is
\begin{equation}
|\psi _{1}\rangle =t_{1}|U\rangle +ir_{1}|L\rangle,
\end{equation}
where $t_{1}$ is amplitude transmissivity of the BS1 and $r_{1}=\sqrt{1-t_{1}^{2}}$ is the corresponding amplitude reflectivity. Here $|L\rangle$ is the photon state reflected by BS1 and going through the lower arm of the outer MZI, while $|U\rangle$ is the photon state propagating along the upper arm of MZI through BS1.

Similarly, after photons passing through mirror E and BS2 of the inner MZI, the state changes to
\begin{eqnarray}
|\psi _{2}\rangle &=&t_{1}e^{i\varphi_{a}}e^{i\phi}|U\rangle +ir_{1}t_{2}e^{i(\varphi_{b}+\varphi_{e})}|u\rangle \nonumber \\
&&-r_{1}r_{2}e^{i(\varphi_{c}+\varphi_{e})}e^{i\chi}|l\rangle,
\end{eqnarray}
where $\varphi_{k}, k=a,b,c,e$ are the tiny complex phases caused by the vibration of mirrors A, B, C and E. State $|l\rangle$ is the photon state reflected by BS2 with amplitude reflectivity $r_{2}$ and propagating along the lower arm of the internal MZI. State $|u\rangle$ is transmitted by BS2 with amplitude transmissivity $t_{2}=\sqrt{1-r_{2}^{2}}$ and propagating along the upper arm of the inner MZI. Due to the phase retarders PR1 and PR2, we introduce two phase shifts $\phi$ and $\chi$ into the state. Here $\phi$ is real phase difference between the two arms of the internal interferometer while $\chi$ is also a real phase difference between the two arms of the outer interferometer.

When photons pass through the beam combiner BS3 in the internal MZI, the state of photons becomes
\begin{eqnarray}
|\psi _{3}\rangle  &=&-(r_{1}t_{2}r_{3}e^{i( \varphi _{e}+\varphi _{b}) }
+r_{1}r_{2}t_{3}e^{i( \varphi _{e}+\varphi _{c}) }e^{i\phi
})|out1\rangle
\nonumber \\
&&-i(r_{1}r_{2}r_{3}e^{i( \varphi _{e}+\varphi _{c})
}e^{i\phi}-r_{1}t_{2}t_{3}e^{i( \varphi _{e}+\varphi _{b}) })|L^{\prime }\rangle  \nonumber\\
&&+t_{1}e^{i\varphi _{a}}|U\rangle,
\end{eqnarray}
where $t_{3}$ and $r_{3}=\sqrt{1-t_{3}^{2}}$ are the amplitude transmissivity and reflectivity of BS3, respectively.  State $|L'\rangle$ refers to the state that photons propagate along the lower arm of external MZI and reach mirror F after the beam combination. While the state $|out1\rangle$ denotes that photons leave NMZI without being detected. 

After BS4, the final photon state from the interferometer is
\begin{equation}
|\psi _{4}\rangle =p_{d}|D\rangle+p_{o1}|out1\rangle+p_{o2}|out2\rangle
\end{equation}
where $p_{d}$, $p_{o1}$, $p_{o2}$ are the probability amplitudes as
\begin{widetext}
\begin{eqnarray}
p_{d}&=&it_{1}r_{4}e^{i\chi}e^{i\varphi_{a}}-ir_{1}t_{4}[r_{2}r_{3}e^{i( \varphi _{e}+\varphi _{c})}e^{i\phi}-t_{2}t_{3}e^{i( \varphi _{e}+\varphi _{b}) }]e^{i\varphi _{f}} \\
p_{o1}&=&-[r_{1}t_{2}r_{3}e^{i( \varphi _{e}+\varphi _{b}) }
+r_{1}r_{2}t_{3}e^{i( \varphi _{e}+\varphi _{c}) }e^{i\phi
}] \label{eqo1}\\ 
p_{o2}&=& t_{1}t_{4}e^{i\varphi _{a}}e^{i\chi }+r_{1}r_{4}[r_{2}r_{3}e^{i(
\varphi _{e}+\varphi _{c}) }e^{i\phi }-t_{2}t_{3}e^{i( \varphi
_{e}+\varphi _{b})}]e^{i\varphi _{f}}.
\label{eqo2}
\end{eqnarray}
\end{widetext}
Here $t_{4}$ and $r_{4}=\sqrt{1-t_{4}^{2}}$ are the amplitude transmissivity and reflectivity of BS4. The state $|D\rangle$ is the photon state finally received by detector D, while another state $|out2\rangle$ means that photons is not detected after passing through NMZI. Then we can obtain the probability of photons reaching the detector as $P_{D} =|p_{d}|^{2}$. Following the three-path interference model\cite{Yuan1, Yuan2}, the complex phases caused by mirror vibrations are usually very small. Under such condition, we can make a linear approximation $e^{i\varphi_{k}}\approx1+i\varphi_k$, where $k=a,b,c,e,f$. Then the probability can be simplified to
\begin{eqnarray}
P_{D} &\approx& -2Im[( \alpha e^{-i\chi }+\beta-\gamma e^{-i\phi }) ( \alpha\varphi
_{a}e^{i\chi }+\beta\varphi _{b}-\gamma\varphi _{c}e^{i\phi }) ] \nonumber \\
&&-2Im
[( \alpha e^{-i\chi }+\beta-\gamma e^{-i\phi })( \beta-\gamma e^{i\phi }) (
\varphi _{e}+\varphi _{f}) ] \nonumber \\
&&+\left\vert \alpha e^{i\chi }+\beta-\gamma e^{i\phi }\right\vert ^{2}.\label{eq9}
\end{eqnarray}
where $\alpha=t_{1}r_{4},\beta=r_{1}t_{2}t_{3}t_{4},\gamma=r_{1}r_{2}r_{3}t_{4}$. Here we only keep the linear terms of complex phase and neglect the high order terms. This is the general output of the NMZI which highly related to the splitting ratios and relative phases $\phi$ and $\chi$. 

\begin{table*}[ht]
\caption{\label{t1}The conditions for eliminating WPI of photons caused by vibrating mirrors. The first column is the WPI to be eliminated while the third column is the corresponding detection method for WPI measurement.}
\begin{ruledtabular}
\begin{tabular}{@{}lll}
WPI to be eliminated&Condition&Detection method\\
\hline
$Im(\varphi_{a})$&$\alpha+\beta cos \chi -\gamma cos ( \phi -\chi )  =0$&position sensitive\\
$Re(\varphi_{a})$&$\beta sin \chi +\gamma sin ( \phi -\chi )=0$&phase sensitive\\
$Im(\varphi_{b})$&$\alpha cos \chi +\beta-\gamma cos \phi=0$&position sensitive\\
$Re(\varphi_{b})$&$\alpha sin \chi -\gamma sin \phi=0$&phase sensitive\\
$Im(\varphi_{c})$&$\alpha cos ( \phi -\chi ) +\beta cos \phi -\gamma=0$&position sensitive\\
$Re(\varphi_{c})$&$\alpha sin ( \phi -\chi ) +\beta sin \phi =0$&phase sensitive\\
$Im(\varphi_{e})$ \& $Im(\varphi_{f})$&$\beta^{2}+\gamma^{2}-2\beta \gamma cos \phi +\alpha \beta cos \chi -\alpha \gamma cos ( \phi -\chi ) =0$&position sensitive\\
$Re(\varphi_{e})$ \& $Re(\varphi_{f})$&$\beta sin \chi +\gamma sin ( \phi -\chi) =0$&phase sensitive\\
\end{tabular}
 \end{ruledtabular}
\end{table*}

The general probability shows that all the WPI caused by vibrating mirrors have reached the detector. And the probability is consist by the complex phases which has two dimensions of information, phase sensitive $Re(\varphi_{k})$ and position sensitive $Im(\varphi_{k})$. Such multiple dimensional informations can be extracted by detection. However, limited by the detection method, the signal we can get is not simply proportional to the probability $P_{D}$. For example, Ref. \cite{Danan} uses a position sensitive quad-cell photodetector with differential detection method to extract the WPI of photons. Such position sensitive detection method can not detect the interference fringes caused by optical phase (real part of complex phases $Re(\varphi_{k}),k=a,b,c,e,f$) but only sensitive to the signal caused by the position, i.e. imaginary part $Im(\varphi_{k})$. Then extracting $Im(\varphi_{k})$ parts in Eq. \ref{eq9} gives the position sensitive signal $i_{pos}$ as
\begin{eqnarray}
i_{pos} &\approx& -2[ \beta^{2}+\gamma^{2}-2\beta\gamma cos\phi+\alpha \beta cos\chi-\alpha \gamma cos(\phi-\chi) ]\nonumber \\
&&\ast[Im(\varphi_{e})+Im(\varphi_{f})]  \nonumber \\
&&-2\alpha[\alpha+\beta cos\chi-\gamma cos(\phi-\chi)]Im(\varphi_{a}) \nonumber \\
&&-2\beta[\alpha cos\chi+\beta-\gamma cos(\phi)]Im(\varphi_{b}) \nonumber \\
&&+2\gamma[\alpha cos(\phi-\chi)+\beta cos\phi-\gamma]Im(\varphi_{c}). \label{eqpo}
\end{eqnarray}

Noticeably, both parts of the complex phases can tell us the WPI. Assuming that the laser beam's waist is large enough, the term with $Im(\varphi_{k})$ can be considered much smaller than the terms with $Re(\varphi_{k})$ (see appendix for detail). Then with a single pixel detector, we can obtain interference fringes only caused by phase terms $Re(\varphi_{k})$. The corresponding phase sensitive signal $i_{pha}$ is
\begin{eqnarray}
i_{pha} &\approx& 2\alpha [\beta sin\chi+\gamma sin(\phi-\chi) ][Re(\varphi_{e})+Re(\varphi_{f})]  \nonumber \\
&&-2\alpha[\beta sin\chi+\gamma sin(\phi-\chi)]Re(\varphi_{a}) \nonumber \\
&&-2\beta[\alpha sin\chi-\gamma sin(\phi)]Re(\varphi_{b}) \nonumber \\
&&+2\gamma[\alpha sin(\phi-\chi)+\beta sin\phi]Re(\varphi_{c}), \label{eqph}
\end{eqnarray}
here we neglect the terms without complex phases. Obviously, the WPI extracted by phase sensitive detection is totally different from the position sensitive one. In the following section, we reconsider the elimination conditions of WPI according to the detection dependent WPI.

\section{Analysis and Discussion}

For simplicity, we start the analysis with the elimination of WPI only caused by mirror A. When using position sensitive detection, the terms with $Im(\varphi_{a})$ in Eq. \ref{eqpo} is $-2\alpha[\alpha+\beta cos\chi-\gamma cos(\phi-\chi)] Im(\varphi_{a})$. Since the vibration of mirror A always causes a position change and the light reflected by mirror A is not zero, we set $Im(\varphi_{a})\neq 0$ and $\alpha\neq 0$. Then to eliminate WPI of mirror A by position sensitive detection, the beam splitting ratios of BSs and phase shifts of NMZI should satisfy
\begin{equation}
\alpha+\beta cos\chi-\gamma cos(\phi-\chi)= 0.  \label{eq12}
\end{equation}
As we can see, the elimination condition of mirror A contains $\alpha$, $\beta$ and $\gamma$ which means that the three paths are interfering simultaneously when using position sensitive detection. But if we use phase sensitive detection, the elimination condition of WPI becomes different. With phase sensitive detection, we can only get the real part of complex phase. Then the WPI caused by mirror A can also be eliminated when
\begin{equation}
\beta sin\chi+\gamma sin(\phi-\chi) = 0.\label{eq13}
\end{equation}
As we can see, there is no $\alpha$ in this condition which is quite different from the position detection case. This is because of the different detection method. Actually, the phase information $Re(\varphi_{a})$ can only be extracted from interference fringes. The absence of $\alpha$ in Eq. \ref{eq13} is induced by a destructive interference of the inner MZI at path $|L'\rangle$ which erasing the WPI from path $|U\rangle$. Then there is no interference fringes can be detected at detector D which erases the WPI of mirror A when using phase sensitive detection. However, this can never happened when using position sensitive method since position induced WPI can be detected even with only one path of light reaching the detector. Therefore, the elimination condition of WPI completely depends on the way of detection. In other words, mirror vibration has multi-dimensional information, which makes WPI have a variety of presentation according to different detection method. 

On the other hand, elimination of WPI of mirror A is due to the interference of multiple paths. In NMZI, interference may occur between any two of the three paths or simultaneously between the three paths. So many kind of interference happening in NMZI results in multiple WPI elimination phenomena. This thinking along, we can further obtain the conditions for eliminating WPI caused by mirrors B, C, E and F from Eq. \ref{eqpo} which shows in table \ref{t1}. As we can see, WPI caused by any mirrors can be eliminated under certain conditions, except erasing all the WPI simultaneously. Since erasing all the WPI simultaneously means no photon can reach detector which is a meaningless result.

Combining the multi-path interference and the detection dependent WPI, WPI elimination effect becomes richer and more controllable than the previous literatures \cite{Vaidman1, Danan, Yuan1}. In the previous work \cite{Vaidman1, Vaidman2, Yuan1}, the WPI caused by mirror E and mirror F can be surprisingly eliminated together. In our case, table 
\ref{t1} shows that WPI caused by any two mirrors can be eliminated at the same time, not only by mirror E and mirror F. What we need to do is carefully adjusting the phase differences and the beam splitting ratios of BSs which satisfy multiple equations in table \ref{t1} at the same time. Even, we can simultaneously eliminate the WPI caused by three mirrors A, E and F according to the table. As we can see in table, elimination conditions of $Re(\varphi_{a})$, $Re(\varphi_{e})$, $Re(\varphi_{f})$ are the same, which means that they can be eliminated together. In this case, the light coming from the laser source can still reach the detector but only contains WPI of mirror B and C. Such phenomenon is not only due to the three-path interference but also the usage of phase sensitive detection method which can be hardly seen in previous work \cite{Vaidman1, Danan, Yuan1, Yuan2, Alonso}. 

\section{Experimental Results}

\begin{figure}[ht!]
\centering
\includegraphics[scale = 0.6]{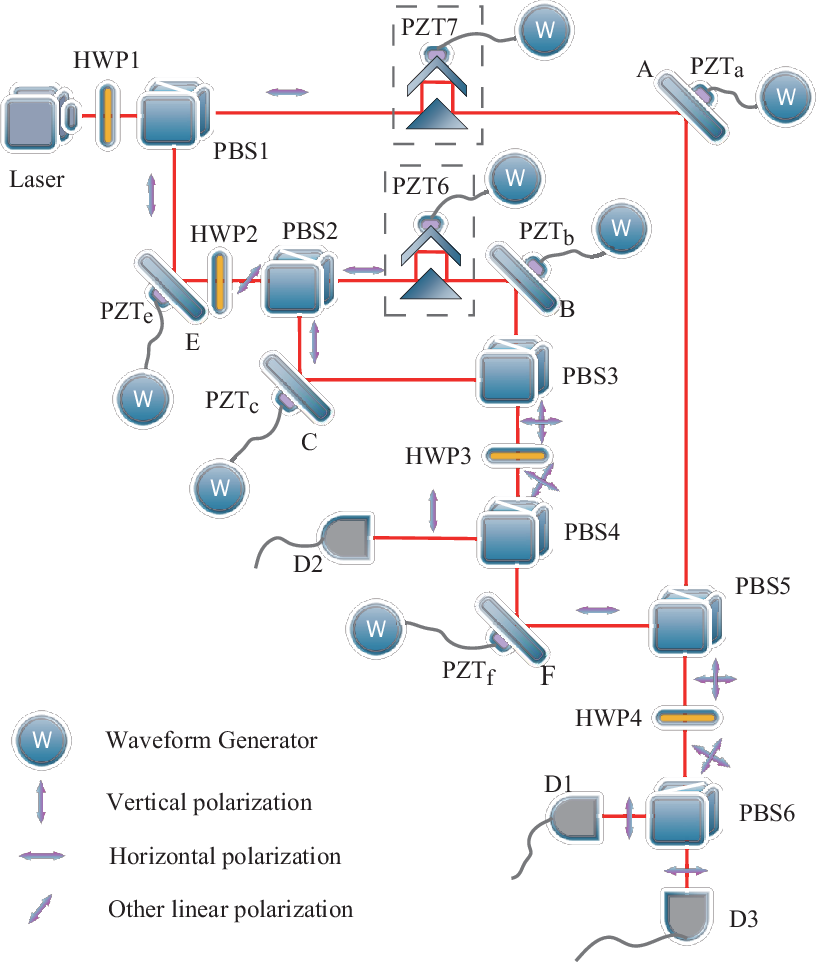}
\caption{ Experimental setup of NMZI. HWP: half wave plate; PBS: polarization beam splitter; D: detectors; PZT: piezoelectric transducer; The lights coming from the laser source passes through the NMZI with different polarizations. Each mirror in the NMZI can vibrate with a PZT who is controlled by a waveform generator. Mirrors A, B, C, E, F are vibrating with $PZT_{k}, k=a,b,c,e,f$ whose frequencies are $f_{a}=1.1kHz$, $f_{b}=1.2kHz$, $f_{c}=1.3kHz$, $f_{e}=1.4kHz$,$f_{f}=1.5kHz$, respectively. The arrows in the figure represents the polarization directions of lights. Inside the two dashed boxes are the phase retarders built by the corner prisms.\label{fig2}}
\end{figure}

To verify the above results, we set up an experiment with phase sensitive detection as shown in Fig. \ref{fig2}. An 1mW 795nm mono color continuous laser from a DFB laser diode is injecting into an interferometer. To avoid detecting the signal caused by imaginary parts of complex phases, we use a Gaussian beam with beam waist 2mm. Such large beam waist ensures that interference fringes we detected are only caused by $Re(\varphi_{k}), k=a,b,c,e,f$, as state in appendix. The light is firstly separated into two arms of the interferometer with different polarization by a half wave plate (HWP1) and a polarization beam splitter (PBS1) as shown in Fig. \ref{fig2}. The combination of half wave plate and PBS acts as an adjustable beam splitter. By changing the angle of HWP1, the laser power in the two arms varies. Then the lights are reflected by the mirrors driving by Piezoelectric Transducers (PZTs) \cite{Danan}. As shown in Fig. \ref{fig2}, each $PZT_{k}, k=a,b,c,e,f$ is driven by a waveform generator which makes each mirror vibrates in different frequencies $f_{k},k=a,b,c,e,f$. 

After passing through the inner interferometer, the laser beams are combined by PBS3 with two orthogonal polarizations as shown in Fig. \ref{fig2}. To make such two orthogonal polarized lights interfere, we use HWP3 to rotate the two lights' polarizations and then use a PBS4 to divide them into horizontal and vertical polarization components. As a result, two beams are mixed in both vertical polarization and horizontal polarizations after PBS4. Such two PBS with a HWP acting as an adjustable beam combiner. The same combination is used in the outer interferometer to get the final output. A single pixel detector D1 is used to detect the output of interferometer. The final output signal of D1 is analyzed by a power spectrum analyzer to obtain the WPI of photons.

According to table \ref{t1}, the two relative phases $\phi$ and $\chi$ play central roles of eliminating the WPI. So the two phases of the NMZI should be accurately controlled and simultaneously detected without affecting WPI. To control the phases $\phi$ and $\chi$, we use corner prisms with PZT6 and PZT7 (as shown in the dashed boxes of Fig. \ref{fig2}) as phase retarders which can adjust the relative optical lengths without changing the visibility of interference fringes. Then we use another two detectors, D2 and D3, to measure the two phases through acquiring the interference signals of the photon states $|out1\rangle$ and $|out2\rangle$. As shown in Fig, \ref{fig2}, D2 is used to measure photon state $|out1\rangle$, the interference signal of inner interferometer. And D3 is used to measure photon state $|out2\rangle$, another output of outer interferometer. Such two interference signals contain the phase information of NMZI when the vibrations of mirror is small enough. According to Eqs. \ref{eqo1}, \ref{eqo2} we can obtain the relation between the two phases $\phi$, $\chi$ and beam splitting ratios $r_{n}, n=1,2,3,4$ through the possibilities of $P_{D2}=|p_{o1}|^{2}$, $P_{D3}=|p_{o2}|^{2}$. Then by neglecting the little vibrations of interference fringes caused by mirrors, we can derive the two phases as
\begin{widetext}
\begin{eqnarray}
\phi  &=&\cos ^{-1}\left[ \frac{P_{D2}-r_{1}^{2}(
t_{2}^{2}r_{3}^{2}+r_{2}^{2}t_{3}^{2}) }{2r_{1}^{2}t_{2}r_{2}t_{3}r_{3}
}\right],   \label{phi}\\
\chi &=&\cos ^{-1}\frac{P_{D3}-(
t_{1}^{2}t_{4}^{2}+r_{1}^{2}r_{2}^{2}r_{3}^{2}r_{4}^{2}+r_{1}^{2}t_{2}^{2}t_{3}^{2}r_{4}^{2}-2r_{1}^{2}r_{2}t_{2}r_{3}t_{3}r_{4}^{2}\cos \phi )
}{2r_{1}t_{1}r_{4}t_{4}\sqrt{%
r_{2}^{2}r_{3}^{2}+t_{2}^{2}t_{3}^{2}-2r_{2}r_{3}t_{2}t_{3}\cos \phi }}%
-\theta,
\end{eqnarray}
\end{widetext}
where $P_{D2}$ and $P_{D3}$ can be obtained with optical powers detected by D2 and D3 which are normalized to the input power, and
\begin{equation}
\theta =\cos ^{-1}\frac{( r_{2}r_{3}\cos \phi -t_{2}t_{3}) }{\sqrt{%
r_{2}^{2}r_{3}^{2}+t_{2}^{2}t_{3}^{2}-2r_{2}r_{3}t_{2}t_{3}\cos \phi }}.\label{beta}
\end{equation}
\begin{figure}[ht!]
\centering
\includegraphics[scale=1]{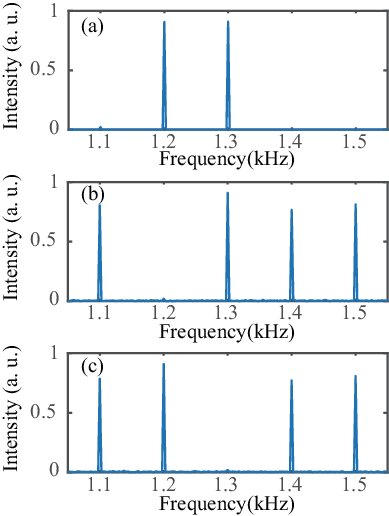}
\caption{The experimental power spectrum of the photon current from D1 shows the elimination of WPI from (a) mirror A, E, F, (b) mirror B, (c) mirror C with phase sensitive detection.\label{fig3}}
\end{figure}

In our experiment, by tuning the angle of HWPs before each PBS, we can easily achieve the adjustment of beam splitting ratios. Then we can adjust the control voltages of PZT6 and PZT7 to change phases $\phi$ and $\chi$. When we adjust the HWPs to set $r_{1}=r_{4}=\sqrt{2/3},r_{2}=r_{3}=\sqrt{1/2}$, the intensity of the light fields in the three paths of NMZI are the same. By varying the two phases $\phi$ and $\chi$, we can observe the elimination of WPI caused by mirror A, B, C and E, F in experiment (as shown in Fig. \ref{fig3}). By using Eqs. \ref{phi}-\ref{beta}, the elimination conditions of WPI caused by mirror A, E and F (as shown in Fig. \ref{fig3}(a)) in experiment are $\phi=0.15rad$ and $\chi=1.67rad$ which is consistent with Eq. \ref{eq13}. When the phases of inner and outer interferometers become $\phi=1.31rad$ and $\chi=0.48rad$ in experiment, the WPI caused by mirror B is eliminated as shown in Fig. \ref{fig3} (b). Moreover, the WPI of mirror C is eliminated as shown in Fig. \ref{fig3} (c) if the two phases are $\phi=0.74 rad,\chi=1.11 rad$. Such experimental results are all consist with the predictions of table \ref{t1}.

\begin{figure}[ht!]
\centering
\includegraphics[scale=0.5]{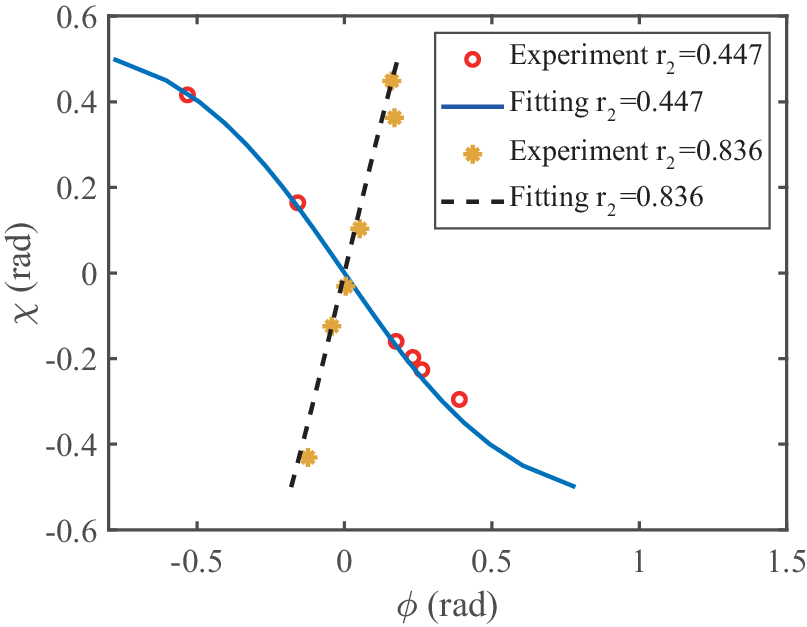}
\caption{The experimental phase conditions to eliminate WPI of mirror A, E and F by phase sensitive detection. The circles are the experimental data and the lines are the corresponding fittings. As we can see, the relation between the two phases $\phi$ and $\chi$ vary with the splitting ratio $r_{2}$ of PBS2 and fit the theoretical predictions. \label{fig4}}
\end{figure}

Furthermore, we experimentally demonstrate the elimination of WPI with different beam splitting ratios in NMZI. Changing the angle of HWP2, the splitting ratio of PBS2 varies. Then to eliminate the WPI of mirror A, we should vary the voltage of PZT6, PZT7 to change phases $\phi$ and $\chi$. As shown in Fig. \ref{fig4}, the red circles are experimental recorded phases $\phi$ and $\chi$ when eliminating WPI of mirror A, E and F at $r_{2}=0.447$. As we can see, the two phases $\phi$ and $\chi$ vary together and fits well with the theoretical simulation (the solid blue line). When the splitting ratio $r_{2}=0.836$, the relation between the two phases changes as shown in yellow circles of Fig. \ref{fig4}. Also, the relation is consistent with the corresponding theoretical simulation (the dashed black line in Fig. \ref{fig4}). These results prove that our theoretical analysis is in good agreement with the experimental facts.

\section{Conclusion}

In summary, we revisit the three-path interference problem which gives the general expression of photons' WPI in NMZI. This general result can not only explain the previous experimental results but also predicted a new WPI elimination phenomenon. More importantly, we propose that there are two kinds of WPI, which respectively relate to the real and imaginary part of complex phase and can be detected by different detection methods. One uses phase sensitive detection which only receive real part related WPI. The other one uses position sensitive detection to measure the imaginary part relating WPI without receiving the real part relating WPI. Due to the different presence of WPI, we discover a new method to eliminate WPI by controlling the beam splitting ratio and phases of interferometer. 

Our results can help to understand the physics of quantum particles. Moreover, such detection dependent WPI elimination phenomenon can be potentially used in quantum information process, especially in quantum cryptography for increasing quantum channels. Of course, thanks to the wide application of NMZI in recent years\cite{Du1, Du2}, some results in this paper may also be applied in quantum metrology. For example, the WPI elimination method is helpful to develop an anti-interference interferometer which can realize the sensing of specific phase information and avoid the disturbing of other stray signals. 

\appendix*
\section{}
To simply demonstrate the phase sensitive detection method, we set the problem in a normal MZI. Considering a Gaussian light with $\Psi(x,y)=A_{0}exp[-(x^{2}+y^{2})/2\Delta^{2}]$ reflected by the vibrating mirror where $A_{0}$ is the amplitude of light and $\Delta$ is the beam waist, the complex phase makes the field become $\Psi(x,y)=A_{0}exp[-(x^{2}+(y+\delta))^{2})/2\Delta^{2}]exp(i\phi)$ where $\delta$ and $\phi$ are the position shift and phase shift caused by mirror vibration. With the tiny vibration assumption, we can only keep the first order term \cite{Danan} which gives the light $\Psi(x,y)=A_{0}exp[-(x^{2}+y^{2})/2\Delta^{2}]exp(-y\delta/2\Delta^{2})exp(i\phi)$. Obviously, the complex phase shift is $\varphi=\phi+iy\delta/2\Delta^{2}$. Then the interference term is
\begin{equation}
2A_{0}^2e^{-\frac{x^{2}+y^{2}}{\Delta^{2}}}e^{-\frac{y\delta}{2\Delta^{2}}}cos\phi\approx 2A_{0}^2e^{-\frac{x^{2}+y^{2}}{\Delta^{2}}}(1-\frac{y\delta}{2\Delta^{2}})cos\phi
\end{equation}
With a large beam waist $\Delta\gg\delta_{a}$, we can neglect the term $y\delta/2\Delta^{2}$ which means the detector can only see the phase sensitive interference fringes $2A_{0}^2exp[-(x^{2}+y^{2})/\Delta^{2}]cos\phi$. That is why a single pixel detector and a large beam waist can achieve the phase sensitive detection. Although we are only discussing a very simple case, the relevant assumptions and conclusions are not difficult to be extended to complex cases, such as in NMZI.

\begin{acknowledgments}
We acknowledge financial support from the National Science Foundation 
of China (Grant NO. 11904227, 12104161, 11804225), the Sailing Program of Shanghai Science and Technology Committee under Grant 19YF1421800, 19YF1414300, Shanghai Municipal Science and Technology Major Project (Grant NO. 2019SHZDZX01), W. Z. also acknowledges additional support
from the Shanghai talent program.
\end{acknowledgments}

\bibliography{apsrev4-2}

\end{document}